%% file: main.tex
\documentclass[sigconf]{acmart}

\acmPrice{}
\makeatletter
\renewcommand\@formatdoi[1]{\ignorespaces}
\makeatother

\usepackage{tikz}
\usepackage{wrapfig}

\AtBeginDocument{%
  \providecommand\BibTeX{{%
    \normalfont B\kern-0.5em{\scshape i\kern-0.25em b}\kern-0.8em\TeX}}}

\setcopyright{none}

\begin{document}

\title{\#TulsaFlop: A Case Study of Algorithmically-Influenced Collective Action on TikTok}


\author{Jack Bandy}
\affiliation{%
  \institution{Northwestern University}
}
\email{jackbandy@u.northwestern.edu}

\author{Nicholas Diakopoulos}
\affiliation{%
  \institution{Northwestern University}
}
\email{nad@northwestern.edu}


\begin{abstract}
When a re-election rally for the U.S. president drew smaller crowds than expected in Tulsa, Oklahoma, many people attributed the low turnout to collective action organized by TikTok users. Motivated by TikTok's surge in popularity and its growing sociopolitical implications, this work explores the role of TikTok's recommender algorithm in amplifying call-to-action videos that promoted collective action against the Tulsa rally. We analyze call-to-action videos from more than 600 TikTok users and compare the visibility (i.e. play count) of these videos with other videos published by the same users. Evidence suggests that Tulsa-related videos generally received more plays, and in some cases the amplification was dramatic. For example, one user's call-to-action video was played over 2 million times, but no other video by the user exceeded 100,000 plays, and the user had fewer than 20,000 followers. Statistical modeling suggests that the increased play count is explained by increased engagement rather than any systematic amplification of call-to-action videos. We conclude by discussing the implications of recommender algorithms amplifying sociopolitical messages, and motivate several promising areas for future work.
\end{abstract}


\maketitle

\input{introduction}

\input{related}

\input{methods}

\input{results}

\input{discussion}



\bibliographystyle{ACM-Reference-Format}
\bibliography{tiktokbib}

\end{document}

%% file: introduction.tex
\section{Introduction}

On June 15th, 2020, the re-election campaign manager for U.S. president Donald Trump shared that an upcoming campaign rally had received more than 1 million ticket requests. The rally was scheduled for the following week in Tulsa, Oklahoma, in a 19,000-seat arena. But fewer than 6,200 people attended the rally, according to the Tulsa Fire Department \cite{restuccia2020trump}. Many have attributed the discrepancy between the campaign's expectations and actual attendance to a protest organized on TikTok, in which users registered for tickets to the rally but did not attend \cite{wells2020tiktok}.

Collective action on social media platforms has been an important research domain in recent years \cite{jackson2020hashtagactivism}, and TikTok presents a distinct new platform for this research area, with crucial differences from other platforms. In particular, while content exposure on Twitter, Facebook, Snapchat, and Instagram primarily depends on social ties determined by the user, TikTok takes a more proactive approach, summed up by one journalist as "[w]hy not just start showing people things and see what they do about it?" \cite{herrman2019tiktok}. Videos on TikTok gain visibility through the algorithmically-curated "For You" feed, which recommends a personalized set of videos to each user -- even if they do not follow anyone in the app \cite{wang2020algorithm}.

Given the centrality of TikTok's recommender algorithm and the platform's growing sociopolitical consequences (as exemplified by the Tulsa rally protest), this work explores how TikTok's recommender algorithm can influence collective action by increasing the visibility of call-to-action videos. To do this, we undertake a focused study of TikTok users who published videos related to the Tulsa rally, aiming to address the following research question: \textit{To what extent did TikTok's recommender algorithm increase the visibility of call-to-action videos related to the Tulsa rally?}

\begin{wrapfigure}{r}{0.45\columnwidth}
    \centering
    \includegraphics[width=0.43\columnwidth]{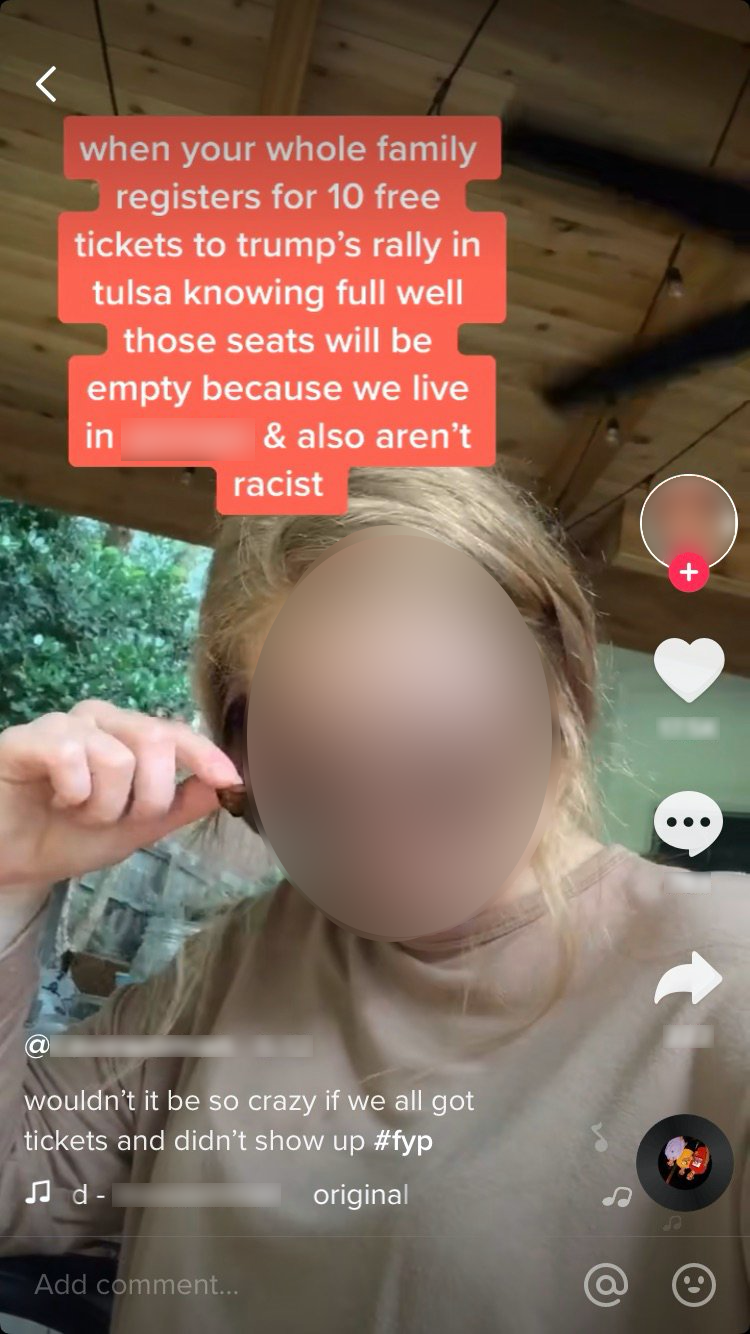}
    \caption{A screenshot from TikTok playing a Tulsa video in our dataset with the caption, "wouldn't it be so crazy if we all got tickets and didn't show up." Some parts of the screenshot are obscured to preserve privacy.}
    \label{fig:screenshot}
\end{wrapfigure}

We begin by collecting a dataset of more than 80,000 videos from over 600 TikTok users who published at least one video related to the Tulsa rally. Our analysis includes overall visibility patterns, user-specific patterns, and statistical modeling, with a focus on how Tulsa-related videos differed from other videos from the same users. Several analyses show that Tulsa-related videos received higher visibility, and some experienced dramatic amplification. For example, one user's call-to-action video was played over 2 million times, but no other video by the user exceeded 100,000 plays, and the user had fewer than 20,000 followers. Statistical modeling suggests that the higher play count for Tulsa videos is associated with higher video engagement (i.e. likes, shares, and comments), rather than any systematic amplification.

We conclude by discussing the challenges of this research area, implications of recommender algorithms amplifying sociopolitical messages, and why the TikTok algorithm is a significant locus of power in today's media ecosystem.

%% file: related.tex
\section{Related Work}
This study intersects with two strains of scholarship: (1) algorithm auditing, an emerging area of research that aims to clarify how algorithmic platforms exercise power in society, and (2) studies that consider social media platforms as sites for collective action, networked counterpublics, and various forms of sociopolitical expression. In the words of Gillespie, we seek to provide "an interrogation of algorithms... with a close attention to where and in what ways the introduction of algorithms into human knowledge practices may have political ramifications" \cite{gillespie2014relevance}.

\subsection{Algorithm Auditing}
As a study measuring the influence of an algorithmic recommender system, this work fits into a large body of algorithm auditing literature. Methods for algorithm audits were first articulated in 2014 by \citet{Diakopoulos2014} and \citet{Sandvig2014auditing}, aiming to achieve a ``public interest scrutiny'' of the powerful yet opaque technologies that affect everyday life: personalization algorithms, search algorithms, recommender algorithms, and more. Since then, a number of algorithm audits have improved the public's understanding of algorithmic information systems: Google image search \cite{Kay2015unequal}, Google web search \cite{Robertson2018google,Trielli2019google}, Apple News \cite{Bandya}, and Facebook ad delivery \cite{Ali2019}, among others.

While algorithm auditing continues to grow as a field, it faces challenges in studying the powerful algorithms used to curate social media feeds. In a 2015 study conducted internally at Facebook, \citet{Bakshy2015facebook} showed how the algorithmically-ranked News Feed slightly decreased exposure to ideologically cross-cutting content, though users' friends and click choices had a stronger influence. Other than this internal Facebook study, algorithm audits of social media platforms have been limited in scale and ecological validity \cite{Lazer2020attention}, especially since platform APIs restrict methods for studying what users actually pay attention to \cite{Bruns2019apicalypse}. One crowdsourced audit measured diversity in the Facebook News Feed  \cite{Bechmann2018newsfeed}, and some audits have focused on other parts of social media platforms such as search (e.g. \cite{Kulshrestha2017bias}). But generally, much remains to be discovered about how algorithms influence exposure on social media, and in particular, little has been explored in relation to TikTok's algorithms.

\subsection{Collective Action on Platforms}
In scholarship related to social media and society, collective action (i.e. people working together to achieve a common objective \cite{britannica}) has been a topic of interest for many researchers. Since the early 2010s, social networks such as Twitter ``have influenced how both those on the margins and those at the center engage in sociopolitical debate and meaning-making,'' as Jackson et al. summarize in their book \textit{\#HashtagActivism: Networks of Race and Gender Justice} \cite{jackson2020hashtagactivism}. For example, \citet{Stewart2017blm} found that Twitter users leveraged the platform affordances in diverse ways when using the \#BlackLivesMatter hashtag. This included asserting group identities and supporting or challenging different frames related to the discourse. Discussing the dynamics of such movements, \citet{Milan2015} uses the term ``cloud protesting'' to describe mobilization and collective action that occurs on social media.

Scholars have also studied collective action and political expression specifically on TikTok. \citet{Literat2019youth} conducted a mixed-methods study of youth-created videos amidst the 2016 US presidential election, finding that young users connected to other users through \textit{content} in the videos, \textit{form} (such as audio choice), and \textit{stance} or ``positioning'' dimensions such as age, gender, sexuality, and race. Through content, form, and stance, users developed shared symbolic resources that enabled connection to ``an assumed like-minded audience'' \cite{Literat2019youth}.

Crucially, algorithmic visibility on these platforms has become key to connecting with others, leading to ``algorithmic entanglements'' where users and algorithms mutually affect one other \cite{gillespie2014relevance}. The concept of algorithmic entanglement is vividly illustrated by Instagram ``pods'' \cite{OMeara2019}, that is, ``groups of users who agree to reciprocally `like' and comment on each other's posts, hoping that increased engagement will boost the odds of appearing on Instagram’s highly trafficked `explore' page'' \cite{petre2019gaming}. Pods represent one type of effort to leverage the platform's algorithm for the users' distribution goals, and the entanglement effect can be observed in Instagram's punitive response to such behavior \cite{petre2019gaming}.

Building on this related work, our study seeks to examine how TikTok's recommender algorithm influenced collective action toward a political rally in the United States. TikTok users registered en masse for a re-election rally held by the U.S. president, then did not attend the rally as an act of protest against the president. Some have called the protest a kind of ``culture jamming'' \cite{wells2020tiktok}, defined in academic literature as ``an interruption, a sabotage, hoax, prank, banditry, or blockage'' that attempts to ``introduce noise into the signal'' \cite{Harold2004pranking}. While the action of registering for tickets was ultimately performed by individuals, TikTok's recommender algorithm also played a significant role in the protest by increasing the visibility of call-to-action videos. Diverging from related work, our study focuses specifically on this role of the algorithm in promoting the videos and thus promoting collective action.

%% file: methods.tex
\section{Methods}

\subsection{Data Collection}
We followed a seed-and-snowball approach to create a dataset for addressing our research question. As a seed, we used a set of seven TikTok videos compiled by journalist Jenna Amatulli \cite{amatulli2020rally}. Each of the videos involved a call-to-action in which the author suggested -- some more explicitly than others -- registering for tickets to the Tulsa rally and then not attending as an act of protest. Some of these videos used hashtags (e.g. \#EmptySeats, \#TulsaFlop) and original soundtracks (also known as ``challenges'') that were subsequently used in other users' call-to-action videos. An open-source TikTok API\footnote{\url{https://github.com/davidteather/TikTok-Api}} allowed us to collect all videos that used these soundtracks and hashtags, snowballing the seed videos into 619 total call-to-action videos, from a total of 616 unique users. We excluded videos published after June 20th, the date of the rally.

We then expanded the dataset to provide a performance baseline with which to compare the call-to-action Tulsa videos.  Using the aforementioned API, we collected up to 500 recent videos from each of the 616 users in our dataset.  Though non-random, this sampling process allowed us to evaluate the Tulsa videos against a baseline of videos from the same users, and expanded the dataset to 80,682 total unique videos. Each video included a unique identifier, a caption, and engagement statistics: like count (how many people have liked the video), comment count (how many comments have been left on the video), share count (how many times users have shared the video, either within the app or to external channels), and play count. Notably, the play count is not explicitly defined by TikTok, though it is generally assumed to reflect the total number of views or "loops" and not the total number of unique users who have viewed it \cite{wang2020algorithm}.

Engagement statistics on TikTok are rounded when greater than one thousand, for example, a video with 5,400,321 likes will only be reported as having 5.4 million likes. This is true for the like count, comment count, share count, and play count for each video, as well as the follower count and following count for each user. In part because of this rounding in the data, which essentially converts metrics into ordinal-like variables (e.g. 5.4 million equals 5.35 million - 5.44 million), we rely heavily on nonparametric measures of rank correlation in our analysis.

\subsection{Inferring and Measuring Algorithmic Visibility}
Our overall goal in analyzing the data was to determine the algorithmic visibility of Tulsa-related videos, that is, the extent to which videos received additional attention through TikTok's recommender algorithm. The effect of the algorithm is impossible to fully isolate given the available data, however, TikTok's disclosures suggest that visibility, engagement, and algorithmic recommendation are closely intertwined in a feedback loop \cite{wang2020algorithm}. Since videos are auto-played as they're recommended to users, we use play count as a proxy for visibility, maintaining that algorithmic recommendation is at least one factor that affects overall visibility. As in prior studies of user-generated video platforms \cite{cha2009analyzing}, which used ``view count'' as a proxy for popularity, this introduces some limitations. For example, a video may have 20 plays, but if two users played it ten times each, then the video's play count is much higher than its actual popularity. Still, due to the relationship between engagement, algorithmic recommendation, and play count on TikTok \cite{wang2020algorithm}, our analysis can reveal the extent to which the recommender algorithm worked in tandem with user behavior to affect the visibility of call-to-action videos in our dataset. Our analytical framework proceeded in three main steps: (1) characterizing overall visibility, (2) measuring user-specific visibility, and (3) modeling visibility.

\subsubsection{Overall Visibility}
We first explored high-level trends in the dataset, focusing on play count as an indicator of visibility. Based on popularity patterns in other user-generated video platforms \cite{cha2009analyzing}, we anticipated a skewed distribution (with most videos receiving limited visibility) in both Tulsa-related videos and other videos in the dataset. We also evaluated the relationship between users' follower counts and their video play counts. One hypothesis about TikTok is that the recommender algorithm will ``just start showing people things and see what they do about it,'' \cite{herrman2019tiktok} rather than relying on follower relationships. To evaluate this hypothesis, we explored potential correlations between users' follower count and their video play counts via Spearman's rank correlation coefficient. This nonparametric measure is sensitive to skewed data, and allowed us to determine which factors correlated with video play count. When comparing the strength of correlations (e.g. whether like count or follower count correlated more strongly with video play count), we used Williams' test of difference between two correlated correlations, due to collinearity in the data.

\subsubsection{User-Specific Visibility}
Since patterns in overall visibility could be attributed to a variety of different factors, we conducted user-centric analysis in the second phase. This allowed us to grasp how algorithmic visibility might affect individual users, and to determine how any overall differences in video visibility played out at the level of individual user experience. Specifically, we determined which users published a Tulsa-related video that received higher visibility compared to other videos they published, as well as the extent of the increase. In this phase, simple measures of central tendency provided sufficient insight for our analysis. We also inspected some individual user accounts manually, and performed the nonparametric paired Wilcoxon signed-rank test (since the data did not meet the normal distribution assumption) to compare users' median videos and Tulsa videos.

\subsubsection{Modeling Visibility}
Finally, we modeled visibility (i.e. play count) as a function of video duration, like count, comment count, share count, author follower count, and whether the video was Tulsa-related. This allowed us to control for other factors affecting the visibility of Tulsa-related videos. Since the play count variable was overdispersed, we used a negative binomial model, and removed videos that received no plays and/or no engagement. Also, since our model used comment count as an independent variable, we excluded videos where the user had disabled comments (0.5\% of the dataset).

The model was intended to determine whether Tulsa videos were associated with a systematic increase in play count when controlling for other factors. We hypothesized that Tulsa-related videos would not experience a systematic increase in play count, expecting that other variables (video duration, like count, etc.) would better explain the variation. The model is only interpreted through this predictive lens, due to multicollinearity among independent variables. For example, since like count and share count are correlated, their coefficients in the model should \textit{not} be interpreted as an explanation of their relative contributions to video visibility on the platform. As we point out in the discussion section, this work lays the foundation for future studies that expand these modeling efforts to better understand visibility dynamics on TikTok.

%% file: results.tex
\section{Results}

\subsection{Overall Visibility}
We first analyzed the overall visibility patterns for the 80,682 videos in our dataset. As expected, the overall distribution of play count was extremely skewed: 86\% of videos had a play count of 1,000 or less, and 37\% had a play count of 100 or less. The play counts generally exhibited a ``long tail'' distribution, as the top 1\% of videos in the dataset claimed 76\% of all plays, and the top 10\% of videos claimed 93\%.

\begin{figure}
    \centering
    \input{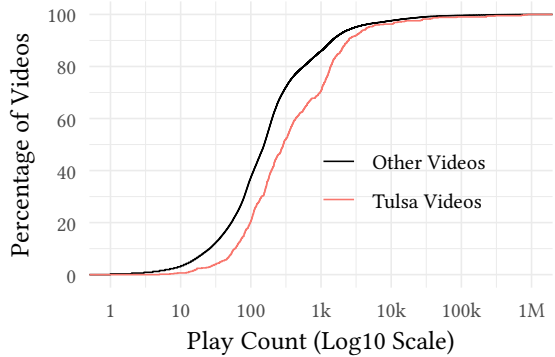}
    \caption{Video play counts are highly skewed (note the logarithmic scale), with 86\% of all videos receiving 1,000 plays or less. Tulsa-related videos received higher play counts.}
    \label{fig:cdfs}
\end{figure}

In a high-level comparison of Tulsa-related videos and other videos, we found Tulsa videos exhibited somewhat higher play counts. For example, 30\% of Tulsa-related videos received more than 1,000 plays, while only 15\% of all other videos received more than 1,000 plays (See Figure \ref{fig:cdfs}). Also, the median Tulsa-related video received 292 plays, while the median video in the rest of the dataset received 155 plays. These surface-level differences suggest potentially higher visibility for Tulsa-related videos, which we explore further in the following analyses. 

\begin{figure}
    \centering
    \input{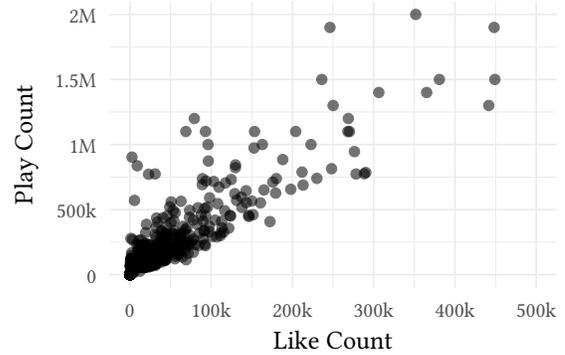}
    \caption{There was a strong correlation ($\rho=0.86$) between a video's play count and the number of users who liked the video, illustrating the algorithmic feedback loop of engagement and amplification, as explained by TikTok \cite{tiktok2020foryou}.}
    \label{fig:likes}
\end{figure}

\begin{figure}
    \centering
    \input{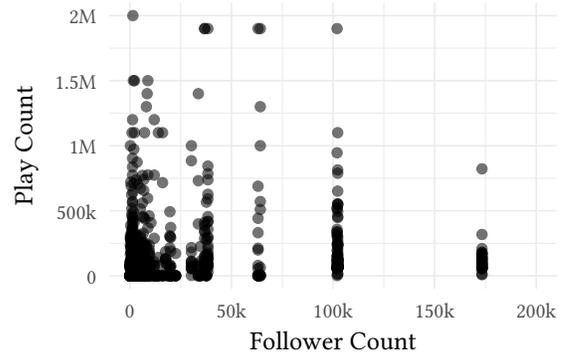}
    \caption{Unlike the strong correlation between play count and like count, there was only moderate correlation ($\rho=0.48$) between a video's play count and the number of users following the author. This supports the notion that videos gain visibility on a case-by-case basis, and that authors' reach is inconsistent.}
    \label{fig:followers}
\end{figure}

In exploring overall video visibility, we found a strong correlation ($\rho=0.86$) between video like count and video play count, but only found a moderate correlation ($\rho=0.48$) between video play count and author follower count (See Figure \ref{fig:likes} and Figure \ref{fig:followers}). Williams' test for difference between two correlated correlations confirmed this difference was statistically significant ($p < 0.001$), thus supporting hypotheses by many users and journalists \cite{wang2020algorithm}. In particular, this evidence supports the notion that videos gain visibility on a case-by-case basis that results mainly from engagement feedback loops (as detailed by TikTok \cite{tiktok2020foryou,wang2020algorithm}), rather than follower relationships. As further evidence, Williams' test confirmed that comment count ($\rho=0.58$) and share count ($\rho=0.52$) also had stronger correlation with play count ($p < 0.001$ for both tests), compared to the correlation between play count and author follower count ($\rho=0.48$).

\subsection{User-Specific Visibility}\label{sec:user-specific}
\begin{figure}
    \centering
    \input{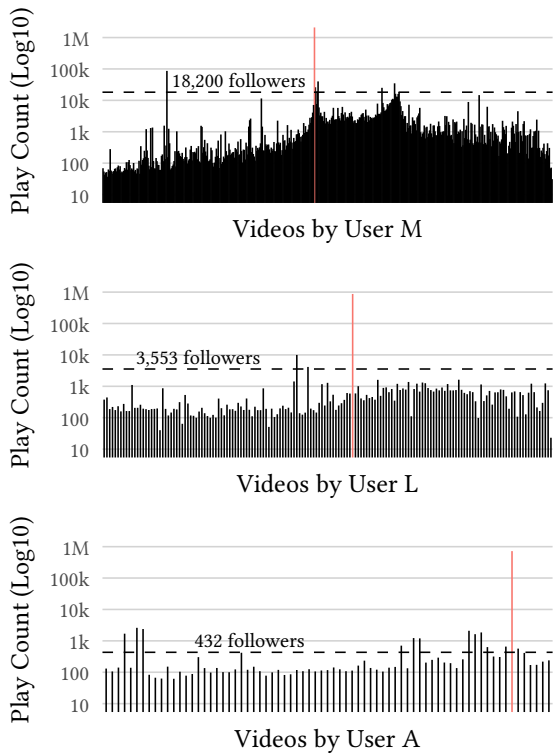}
    \caption{
    Three users whose call-to-action videos (red, and highest bar in each of the three charts) received a dramatic increase in visibility. For example, User M's call-to-action video was played over 2 million times, but no other video exceeded 100,000 plays, and the user had fewer than 20,000 followers as of early July. Usernames are kept anonymous to preserve privacy.}
    \label{fig:user_bar_plots}
\end{figure}

To understand how algorithmic visibility played out for individual users, we analyzed how users' Tulsa-related videos compared to other videos they published. For 9\% of users, their Tulsa-related video was their most-played video. The three greatest benefactors are shown in Figure \ref{fig:user_bar_plots}, which depicts their Tulsa videos experiencing a dramatic increase in play count compared to their other videos and their follower count.

\begin{figure}
    \centering
    \input{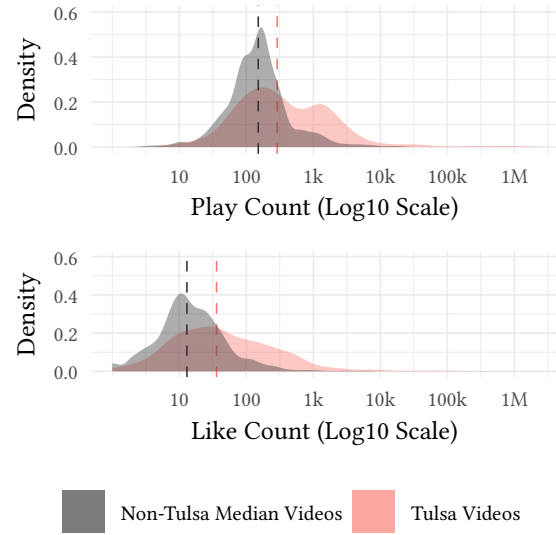}
    \caption{
    Compared to users' median videos (gray), users' Tulsa-related videos (red) were associated with higher play counts in our dataset. The dashed line represents the median user.}
    \label{fig:distribution}
\end{figure}

While not all users' Tulsa videos experienced a dramatic visibility increase (like the users in Figure \ref{fig:user_bar_plots}), 76\% of users did experience an increase of some kind compared to their median video. These gains were highly skewed, with an average increase of 9,606 additional plays and the median user experiencing an increase of 90 additional plays (IQR: 2 - 885 additional plays) (See Figure \ref{fig:distribution}). Furthermore, most users' Tulsa videos also received more \textit{likes} than their median video. This effect was also skewed, with an average increase of 2,274 additional likes and the median user experiencing an increase of 14 likes (IQR: 0 - 96 additional likes). Wilcoxon signed rank tests suggest the shifts were statistically significant for play count, like count, share count, and comment count ($p < 0.001$ in all four tests). Figure \ref{fig:distribution} visualizes these differences.

\subsection{Modeling Visibility}

Our negative binomial regression model found no statistically significant increase in play count associated with Tulsa-related videos. In other words, after controlling for engagement variables and other factors captured in the dataset, call-to-action Tulsa videos did not receive uniquely higher play counts relative to non-Tulsa videos. Instead, the higher play counts in our dataset are associated with factors including video length, like count, comment count, share count, and author follower count. These factors were statistically significant ($p < 0.001$) in the model, as detailed in Table \ref{table:coefficients}.

\begin{table}
\begin{center}
\begin{tabular}{l c}
\hline
 & Incident rate ratio \\
\hline
Video Length     & $0.98890^{***}$ \\
Like Count          & $1.00012^{***}$  \\
Comment Count       & $1.00212^{***}$  \\
Share Count         & $0.99967^{***}$ \\
Follower Count      & $1.00001^{***}$  \\
Tulsa Video      & $1.00633$        \\
\hline
Pseudo-R² (Cragg-Uhler) & $0.66949$   \\
\hline
\multicolumn{2}{l}{\scriptsize{$^{***}p<0.001$; $^{**}p<0.01$; $^{*}p<0.05$}}
\end{tabular}
\caption{Incident rate ratios for the independent variables in the negative binomial model predicting video play count. For example, if a video were one second longer, the model predicts a decrease in play count by a factor of 0.98890, holding all other variables constant. On the other hand, if a video receives one additional like, the model predicts an increase in play count by a factor of 1.00012, holding all other variables constant.}
\label{table:coefficients}
\end{center}
\end{table}

While Tulsa Videos were associated with an increase in play count, the model did not find that factor statistically significant, and the other independent variables better explained the variation. The incident rate ratios for these variables should only be interpreted as predictive of play count within our dataset, as they do not account for many other signals that TikTok uses (e.g. language preference, device type, etc. \cite{tiktok2020foryou}). For example, the model found a \textit{negative} association between share count and play count, which contradicts TikTok's characterization that sharing a video is a positive signal to the algorithm. At the same time, some of the associations align with TikTok's characterizations: favorable responses such as "likes" and "comments" will allegedly signal the algorithm to recommend the video to more users, and indeed, the model associates more likes and comments with an increase in play count. Also, longer videos are intuitively associated with a decrease in play count, as shorter videos are more likely to be ``looped'' and receive additional plays.


%% file: discussion.tex
\section{Discussion}

\subsection{Limitations}
This case study showed how TikTok's algorithmic recommender system can influence collective action by working alongside user behavior to increase the visibility of some videos. This work suffered from several notable limitations, especially as one of the first research efforts exploring TikTok's recommender algorithm. First, the corpus of videos was generated through a non-random snowball sample, and likely excluded other relevant videos. The video statistics were also collected more than one week after the Tulsa rally, and some statistics could have changed following media attention after the event (i.e. \cite{wells2020tiktok}). Also, some metrics we used in the paper were rounded, which is an inherent limitation of TikTok data as of writing.

While we present evidence that some users posted a call-to-action video that was amplified by TikTok's recommender algorithm, the most important limitation of our study is the reliance on play count as a proxy for algorithmic visibility. For example, the users in Figure \ref{fig:user_bar_plots} published a call-to-action video that received dramatically more plays than their second most-played video by an order of magnitude, however, it remains unknown how many of these additional plays came from the algorithmic ``For You'' recommender feed, how many came from ``organic'' views of the video (i.e. by users directly visiting the profile), and how many came from other areas of the app such as the trending ``Discover'' tab--another algorithmic recommender feed. The inability to fully isolate the effect of the ``For You'' algorithm is an inherent but significant limitation of our study.

\subsection{Power to the People or Power to the Algorithm?}
This case study of algorithmically-influenced collective action highlights the power dynamics in platform technologies. Langdon Winner posited two main forms in which technology affects power dynamics \cite{winner1980artifacts}: (1) politically \textit{arranged} artifacts and (2) \textit{inherently} political artifacts. In Winner's examples, a highway is a politically \textit{arranged} technology when designed to exclude buses and public transportation options used by the lower class, whereas the atomic bomb is an \textit{inherently} political technology since it essentially requires a hierarchical power structure for safe management. TikTok's platform, driven mainly by algorithmic recommendations, embodies aspects of both inherently political technologies as well as politically arranged technologies.

As an inherently political technology, Langdon Winner and Lewis Mumford might have described TikTok as democratized rather than ``authoritarian,'' because the platform is largely user-driven. In Mumford's words, democratic technologies are ``man-centered, relatively weak, but resourceful and durable'' \cite{mumford1964authoritarian}. User-driven amplification, as evidenced in our analyses, is an important democratic aspect of TikTok which allows anyone the chance to "go viral," in a sense: if enough users ``respond favorably'' to a video (i.e. liking or commenting), the algorithm may recommend it to a much larger audience of users \cite{wang2020algorithm}, regardless of who follows the author. This is in addition to TikTok's laissez-faire approach to permitted content -- as one spokeswoman summarized, users ``can post videos on whatever is interesting and expressive to them, aside from things like hate speech'' \cite{wells2020politics}. Such an approach means that users can (and likely will) continue using the platform as "an information and organizing hub" \cite{lorenz2020ban} for grassroots political organization similar to the kind featured in this study.

But some aspects of TikTok push the platform toward Mumford's category of ``authoritarian'' technologies, described as ``system-centered, immensely powerful, but inherently unstable'' \cite{mumford1964authoritarian}. Namely, users have limited control over what becomes visible on the platform, and visibility is highly skewed -- the top 1\% of videos in our dataset accounted for 76\% of all plays. While the company has revealed some aspects of its algorithm and promises to reveal more \cite{wang2020algorithm}, much remains unknown about its algorithmic practices. Even if the company achieved full transparency, it would not automatically empower users to gain visibility. Furthermore, social media platforms have sometimes taken countermeasures to squelch organizational efforts, such as ``Instagram pods,'' which attempt to improve algorithmic visibility through coordinated engagement \cite{petre2019gaming}. We noticed coordinated engagement efforts in some TikTok videos (e.g. ``hit share, copy link to help promote the video''), however, little is known about any countermeasures from TikTok and how they may affect visibility. These countermeasures, along with TikTok's content moderation practices, advertising policies, and enforcement of community guidelines,\footnote{\url{https://www.tiktok.com/community-guidelines}} could potentially turn TikTok into a more ``authoritarian'' technology with central points of control.

\subsection{Future Work}
As one of the first studies exploring algorithmic media exposure on TikTok, this work surfaces many promising areas for future research. Subsequent studies might collect data at a larger scale by combining our seed-and-snowball approach with other sampling methods. A truly random sample is nearly impossible given TikTok's vast scale and restricted API, but a larger pool of data could afford a pseudo-random sample to serve as a compelling baseline for studying other groups of videos and users on the platform.

In addition to video visibility patterns and algorithmic amplification, many other aspects of TikTok also deserve research attention. Future audits may explore how the recommendation algorithm progresses from its ``warm start,'' where users select broad categories (e.g. cooking, dancing, animals, sports, etc.), to highly-personalized recommendations in the ``For You'' feed. Studying algorithms in other sections of the app (such as ``Following,'' ``Trending,'' and ``Discover'') may help illuminate how some videos and users gain visibility and popularity. The account suggestion algorithm should also be explored, as early evidence suggests significant racial bias \cite{Strapagiel2020bias}.

Future studies of the platform may especially benefit from qualitative research methods, which could potentially surface important platform dynamics not captured in quantitative analyses. For example, some videos and captions encouraged users to take certain steps ``for the algorithm,'' such as liking and commenting on the video to increase its perceived engagement. Also, one video we viewed \textit{redacted} an original call-to-action video because the author thought registering for the rally would ``boost the president's ego'' and provide the re-election campaign with the protesters' personal data. This kind of activity brings up a number of interesting dynamics for future research, especially when combined with the complexity of algorithmic visibility.

Finally, future work should also seek to further isolate algorithmic visibility. This would primarily entail statistical or machine-learned models that predict visibility from engagement metrics and other video features. Building on the modeling in this work, future models may include features such as the time and date when a video was posted, sounds used in the video, hashtags in video captions, users mentioned, and more.  For researchers who want to more fully understand the role of recommender algorithms in the media ecosystem, attributing visibility to algorithmic amplification within complex sociotechnical assemblages is a high-priority methodological challenge. Otherwise, the ``interrogation of algorithms'' \cite{gillespie2014relevance} risks being only an ``interrogation of platforms,`` and may not appropriately account for the role of algorithms in modern sociopolitical activity.